\documentclass[aps,prl,reprint,twocolumn,amsmath,amssymb,showpacs,superscriptaddress, longbibliography]{revtex4-1}

\usepackage{color}
\usepackage{textcomp} 
\usepackage{mathrsfs,amsmath}
\usepackage{graphicx}
\usepackage{dcolumn}
\usepackage[mathlines]{lineno}
\usepackage{hyperref}
\usepackage{xcolor}
\hypersetup{
    colorlinks,
    linkcolor={red!50!black},
    citecolor={blue!50!black},
    urlcolor={blue!80!black}
}
\usepackage{bm}
\usepackage{epstopdf}
\usepackage{soul}
\usepackage{epsfig}
\usepackage{bbold}
\usepackage{braket}
\usepackage{physics}
\usepackage{gensymb}
\usepackage{amsmath,amssymb}

\begin{document}

\title{Comment on M. Babiker, J. Yuan, K. Koksal, and V. Lembessis, Optics Communications 554, 130185 (2024)}
\author{Kayn A. Forbes}
\email{K.Forbes@uea.ac.uk}

\affiliation{School of Chemistry, University of East Anglia, Norwich Research Park, Norwich NR4 7TJ, United Kingdom}

\begin{abstract}
    In a recent article Babiker et al. [Optics
Communications $\mathbf{554}$, 130185 (2024)] claim that cylindrical vector beams (CVBs), also referred to as higher-order Poincar\'e  (HOP) beams, possess optical chirality densities which exhibit `superchirality'. Here we show that, on the contrary, CVBs possess less optical chirality density than a corresponding circularly polarized scalar vortex beam and that the `superchiral' results are nonphysical. We also identify a number of issues concerning the derivation and general theory presented. 
\end{abstract}

\maketitle

\section{Transverse fields}

Babiker et al. make claims in \cite{babiker2024super} (referred to as BYKL from now on) of `superchirality' (they also refer to it as `superhelicity') in cylindrical vector beams (CVBs), also called higher-order Poincar\'e  beams (HOPs). The results of BYKL, however, appear incorrect. In the first part of this Comment we look purely at the transverse polarized electromagnetic fields in BYKL, we comment on the longitudinal $z$-polarized components later. Combining these results we produce the correct form of Eq. (15) of BYKL. We then highlight that the claimed `superchirality' is nonphysical. We conclude by raising more general issues with the theoretical methods and physics presented in BYKL. 

The vector potential $\textbf{A}$ for a circularly polarized scalar vortex beam in cylindrical coordinates $(r,\phi,z)$ is \cite{gbur2016singular, zhan2009cylindrical}

\begin{align}
\mathbf{A} &= (\mathbf{\hat{\boldsymbol{\rho}}}\pm i \mathbf{\hat{\boldsymbol{\phi}}})F(r, z)\text{e}^{i((m\pm1)\phi+k_z z - \omega t)},
\label{eq:1}
\end{align}

where the upper sign of $\pm$ corresponds to left-handed circular polarization and the lower sign to right-handed circular polarization; $\mathbf{\hat{\boldsymbol{\rho}}}$ and $\mathbf{\hat{\boldsymbol{\phi}}}$ are the radial and azimuthal unit vectors for cylindrical coordinates, respectively; $F(r, z)$ is the amplitude distribution; $m \in \mathbb{Z}$ is the topological charge; $k_z$ is the longitudinal wavenumber; $\omega = ck$ is the angular frequency. The electric field is derived by $\mathbf{E}=-\frac{\partial \textbf{A}}{\partial t}$ (note BYKL conflate $k$ and $k_z$):

\begin{align}
\mathbf{E} &= i\omega(\mathbf{\hat{\boldsymbol{\rho}}}\pm i \mathbf{\hat{\boldsymbol{\phi}}})F(r, z)\text{e}^{i((m\pm1)\phi+k_z z - \omega t)}.
\label{eq:2}
\end{align}

The magnetic field is derived by taking the curl of the vector potential, $\mathbf{B}= \nabla \times \mathbf{A}$ (we only require the derivative with respect to $z$ as we are looking purely at the transverse field components at this point)

\begin{align}
\mathbf{B} &= \frac{d}{dz}\mathbf{\hat{z}} \times (\mathbf{\hat{{\boldsymbol{\rho}}}}\pm i \mathbf{\hat{\boldsymbol{\phi}}})F(r, z)\text{e}^{i((m\pm1)\phi+k_z z - \omega t)} \nonumber \\
&  = i k_z (\mathbf{\hat{\boldsymbol{\phi}}}\mp i \mathbf{\hat{\boldsymbol{\rho}}})F(r, z)\text{e}^{i((m\pm1)\phi+k_z z - \omega t)} \nonumber \\
& = k_z (i\mathbf{\hat{\boldsymbol{\phi}}}\pm  \mathbf{\hat{\boldsymbol{\rho}}})F(r, z)\text{e}^{i((m\pm1)\phi+k_z z - \omega t)} 
\label{eq:3}
\end{align}

In the above, following BYKL, the $z$-derivative of the amplitude distribution is assumed small and ignored. Using Eqs.~\eqref{eq:2} and \eqref{eq:3} a left-handed circularly polarized beam has the following transverse electromagnetic fields: 

\begin{align}
\mathbf{E_1} &= i\omega(\mathbf{\hat{\boldsymbol{\rho}}} + i \mathbf{\hat{\boldsymbol{\phi}}})\sin(\Theta/2) F_1(r, z)\text{e}^{i((m+1)\phi+k_z z - \omega t)},
\label{eq:4}
\end{align}

\begin{align}
\mathbf{B_1} &=  k_z (i\mathbf{\hat{\boldsymbol{\phi}}}+  \mathbf{\hat{\boldsymbol{\rho}}})\sin(\Theta/2) F_1(r, z)\text{e}^{i((m+1)\phi+k_z z  - \omega t)}, 
\label{eq:5}
\end{align}

which agree with the transverse components of Eqs.~(10) and (11) in BYKL, giving the correct optical chirality density. i.e. $\eta \propto -\text{Im}(\mathbf{E_1}^* \cdot \mathbf{B_1}) \propto |F_1|^2\sin^2(\Theta/2)  $ (where the superscript * denotes complex conjugate). We have introduced the Poincar\'e angle $\Theta$ (as defined in BYKL) to allow us to vary the latitude of an arbitrary point on a higher-order Poincar\'e sphere. We don't need to account for longitude Poincar\'e angle because under the conditions used throughout BYKL it does not influence the optical chirality (see Further Comments below for more discussion). However, for the right-handed circularly-polarized beam: 

\begin{align}
\mathbf{E_2} &= i\omega(\mathbf{\hat{\boldsymbol{\rho}}} - i \mathbf{\hat{\boldsymbol{\phi}}})\cos(\Theta/2) F_2(r, z)\text{e}^{i((m-1)\phi+k_z z  - \omega t)},
\label{eq:6}
\end{align}

\begin{align}
\mathbf{B_2} &=  k_z (i\mathbf{\hat{\boldsymbol{\phi}}} -  \mathbf{\hat{\boldsymbol{\rho}}})\cos(\Theta/2) F_2(r, z)\text{e}^{i((m-1)\phi+k_z z  - \omega t)}.
\label{eq:7}
\end{align}

Eq.~\eqref{eq:7} does not match the transverse components of Eq.~(12) in BYKL. Inserting Eqs.~\eqref{eq:6} and \eqref{eq:7} into $\eta  \propto -\text{Im}(\mathbf{E}^* \cdot \mathbf{B})$ gives the optical chirality density for a right-handed polarization state with the correct sign: $\eta  \propto -\text{Im}(\mathbf{E_2}^* \cdot \mathbf{B_2}) \propto -|F_2|^2 \cos^2(\Theta/2)$. The optical chirality of these two beams, when added together give $\eta  \propto |F_1|^2\sin^2 (\Theta/2)- |F_2|^2\cos^2 (\Theta/2)$ to represent a CVB. If the beams have equal amplitude ($F_1=F_2=F$, as in BYKL) then $\eta  \propto -\cos (\Theta) |F|^2$. In contrast, the first term in Eq.~ (15) of BYKL is $\eta  \propto 2\cos (\Theta) |F|^2$ and gives the wrong sign for the optical chirality. 

\section{Longitudinal fields}
BYKL also presents the longitudinal $z$-polarized electric and magnetic field components of a CVB. For a general circularly polarized scalar optical vortex these are as follows \cite{quinteiro2019reexamination, forbes2021relevance}:

\begin{align}
E_z &= -\frac{\omega}{k_z}\Bigl(\frac{\partial}{\partial r}-\frac{\sigma m}{r}\Bigr) F(r,z) \text{e}^{i[(m \pm 1)\phi+{k_z}z]},
\label{eq:8}
\end{align}

and, 

\begin{align}
B_z &= i\Bigl(\sigma \frac{\partial}{\partial r}-\frac{ m}{r}\Bigr) F(r,z) \text{e}^{i[(m \pm 1)\phi+{k_z}z]},
\label{eq:9}
\end{align}

where $\sigma = \pm 1$ is the helicity, the positive sign corresponding to left-circular polarization and the lower sign corresponding to right-circular polarization. A CVB or HOP beam consists of a superposition of a left-handed circularly polarized vortex beam with a right-handed vortex charge $m<0$ and the right-circularly polarized component has a left-handed vortex charge $m>0$ \cite{rosales2018review}. Using Eqs.~\eqref{eq:2}, \eqref{eq:3}, \eqref{eq:8}, and \eqref{eq:9}   we can now write the fully correct form of equations (10)-(13) of BYKL. (Note, importantly, here we label the topological charge of beam 1 as $m_1$ and of beam 2 as $m_2$ in order to clearly highlight the issues in BYKL):

\begin{align}
\mathbf{E_1} &= \omega\Bigl[i(\mathbf{\hat{\boldsymbol{\rho}}} + i \mathbf{\hat{\boldsymbol{\phi}}}) - \frac{\mathbf{\hat{z}}}{k_z}\Bigl(\frac{\partial}{\partial r}+\frac{|m_1|}{r}\Bigr)\Bigr] F_1(r, z) \nonumber \\
& \times \text{e}^{i[(-|m_1|+1)\phi+k_z z)]}\sin(\Theta/2)
\label{eq:10},
\end{align}

\begin{align}
\mathbf{B_1} &=  \Bigl[k_z(i\mathbf{\hat{\boldsymbol{\phi}}} +  \mathbf{\hat{\boldsymbol{\rho}}}) + i\mathbf{\hat{z}}\Bigl(\frac{\partial}{\partial r}+\frac{|m_1|}{r}\Bigr)\Bigr] F_1(r, z) \nonumber \\
& \times \text{e}^{i[(-|m_1|+1)\phi+k_z z)]}\sin(\Theta/2)
\label{eq:11},
\end{align}

\begin{align}
\mathbf{E_2} &= \omega\Bigl[i(\mathbf{\hat{\boldsymbol{\rho}}} - i \mathbf{\hat{\boldsymbol{\phi}}}) - \frac{\mathbf{\hat{z}}}{k_z}\Bigl(\frac{\partial}{\partial r}+\frac{|m_2|}{r}\Bigr)\Bigr] F_2(r, z) \nonumber \\
& \times \text{e}^{i[(|m_2|-1)\phi+k_z z)]}\cos(\Theta/2)
\label{eq:12},
\end{align}

\begin{align}
\mathbf{B_2} &=  \Bigl[k_z(i\mathbf{\hat{\boldsymbol{\phi}}} -  \mathbf{\hat{\boldsymbol{\rho}}}) + i\mathbf{\hat{z}}\Bigl(-\frac{\partial}{\partial r}-\frac{|m_2|}{r}\Bigr)\Bigr] F_2(r, z) \nonumber \\
& \times \text{e}^{i[(|m_2|-1)\phi+k_z z)]}\cos(\Theta/2)
\label{eq:13}.
\end{align}

Using Eqs.~\eqref{eq:10}-\eqref{eq:13} to calculate the optical chirality density gives:

\begin{align}
\eta  & = \frac{\epsilon_0}{4k} \Big[\omega k_z (|F_{1}|^2\sin^2(\Theta/2)-|F_{2}|^2\cos^2(\Theta/2)) \nonumber \\ &+\frac{\omega}{k_z}\Bigl\{\sin^2(\Theta/2)\Bigl( |{F_1^{'}}|^2 + \frac{2 |m_1| {F_1^{*'}}F_1}{r} + \frac{|m_1|^2|F_1|^2}{r^2} \nonumber \\ & - \cos^2(\Theta/2)\Bigl(|{F_2^{'}}|^2 + \frac{2 |m_2|{F_2^{*'}}F_2}{r} +  \frac{|m_2|^2|F_2|^2}{r^2} \Bigr)\Bigr\} \Bigr], 
\label{eq:14}
\end{align}

where the superscript prime denotes differentiation with respect to $r$. This correct result for the optical chirality also matches exactly that of Eq.~(14) in \cite{Forbes2024optical} when the second-order transverse fields are neglected as in BYKL (see Further Comments for more discussion on this). BYKL explicitly studies a CVB or HOP beam, and so the topological charges $m_1$ and $m_2$ of the two beams in the superposition are equal in magnitude $|m_1| = |m_2|$ by definition \cite{rosales2018review}. BYKL therefore use the label $m$. This is problematic notation because $m$ can take both positive and negative values. Boundary conditions were put upon $m$ at the very start of BYKL which studied a superposition of circularly polarized scalar vortices with opposite topological charge handedness to polarization handedness (this is why we have used the modulus in the appropriate places in this Comment). Remember CVBs are just a small subset of cylindrical vector vortex beams (CVVBs) \cite{rosales2018review, yi2015hybrid, arora2020detection, arora2020hybrid}. Nonetheless, 
noting also that $F_1=F_2=F$ in BYKL we can simplify further to get:    

\begin{align}
\eta  & = -\cos\Theta\frac{\epsilon_0}{4k} \Big[\omega k_z |F|^2 +\frac{\omega}{k_z}\Bigl( |{F^{'}}|^2\nonumber\\  & + \frac{2 |m| {F^{*'}}F}{r} + \frac{|m|^2|F|^2}{r^2} \Bigr) \Bigr],
\label{eq:15}
\end{align}

which has a different form compared to Eq. (15) in BYKL. 

\section{`superchirality'}

The simulations (Figure 2 in BYKL) claiming to show `superchirality' of CVBs based on Eq. (15) of that paper are misleading. Firstly it is important to point out that the simulations given in BYKL correspond to the case of where $\cos \Theta = 0$. This corresponds to the north pole on any higher-order Poincar\'e sphere, which is in fact just a right-circularly polarized scalar mode of vortex charge $|m|$. Thus the optical chirality distribution given in Figure 2 of BYKL is that of a \textit{scalar} vortex beam with a homogenous spatial distribution of right-circular polarization (though note the sign is incorrect in BYKL), not a cylindrical \textit{vector} beam. A CVB would manifest when $0 < \Theta < \pi$. This clearly shows that, all other parameters being equal, as you move from the poles of any higher-order Poincar\'e sphere the CVB possesses less optical chirality than a scalar vortex beam represented on the poles, eventually reaching zero on the equator $\Theta = \pi/2$. Equally important with regards to `superchirality' is the fact that it appears the long-established upper-bound on topological charge density of an optical vortex beam \cite{roux2003optical} has not been considered in BYKL. No mention is given of the beam waist $w_0$ or wavelength $\lambda$ used in the simulation. Nonetheless, if we first assume a very tight focus, $w_0$ = $\lambda$ and $\lambda = 729~\text{nm}$ then according to \cite{roux2003optical}: $|m|/2\pi w_0 < 1/\lambda$ which is easily rearranged to show that for such a beam waist $|m| < 2\pi$. Given these parameters, we present in Figure 1 the optical chirality density using the well-established formula for Laguerre-Gaussian beams \cite{green2023optical, forbes2021measures}. 

Note, importantly, that the optical chirality density shows no `superchiral' behaviour, and the peak value for any given $|m|$ decreases with increasing $|m|$ due to the spreading out of the spatial distribution and decreasing density. We attempt to reproduce Figure 2 of BYKL in Fig.~\ref{fig:2} of this Comment. In order to emulate the (nonphysical) results of BYKL the beam waist was required to be a not experimentally achievable $w_0 \approx \lambda / 5$ for $\lambda = 729~\text{nm}$. However, taking into account the bound on the topological charge density, the allowed values of $|m|$ for a beam with a circumference of $2\pi \lambda /5$ is $|m| < 2$. This therefore means that the 
behaviour exhibited in Figure 2 of BYKL for any value of $|m| >1 $ is nonphysical and for the `allowed' values of $|m|=0,1$ the required beam waist is experimentally prohibited. 

\begin{figure*}
    \includegraphics[]{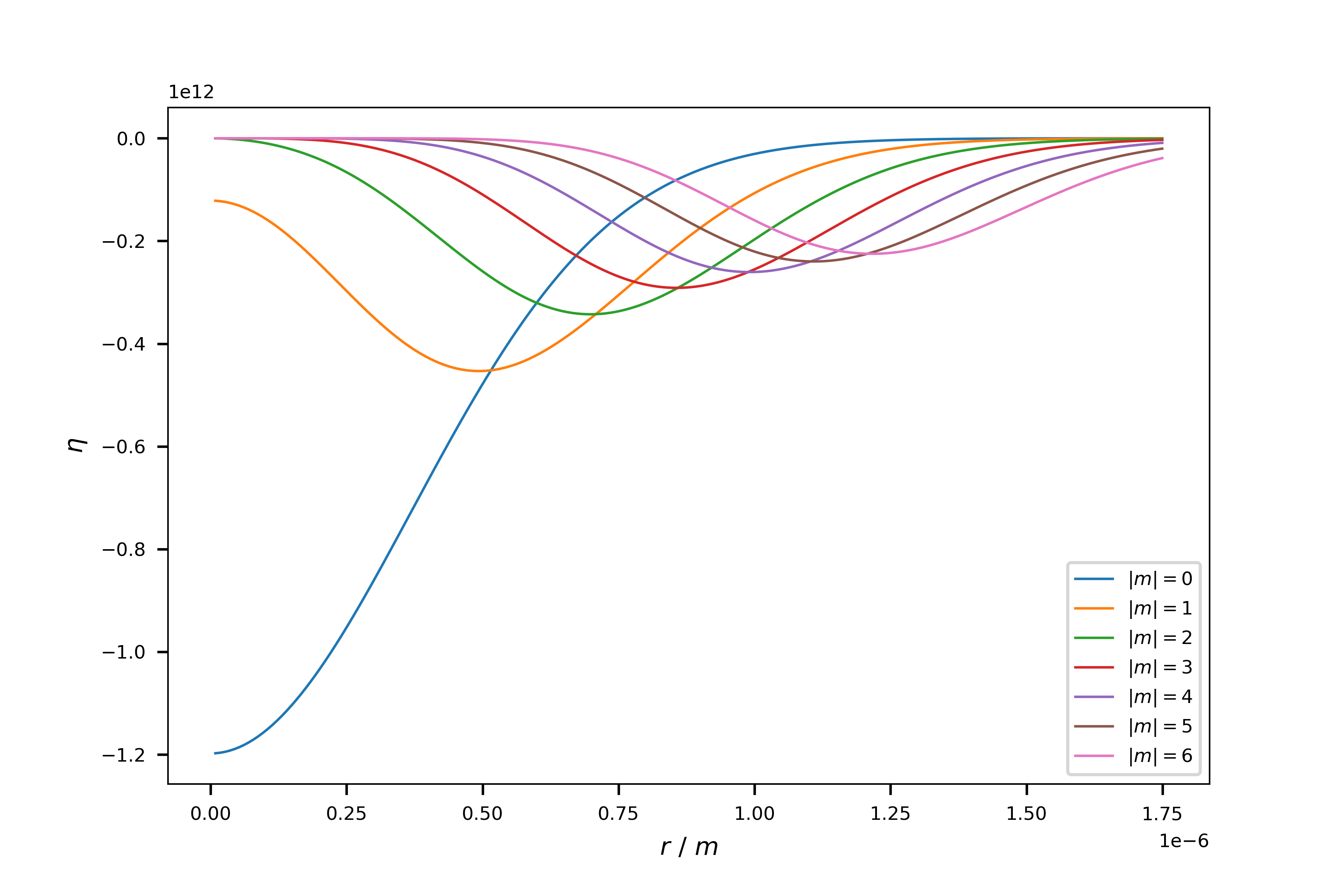}
    \caption{Optical chirality density \eqref{eq:15} for a Laguerre-Gaussian mode $(m,p=0)$ and $\cos \Theta = 0$, i.e. a right-circularly polarized scalar Laguerre-Gaussian vortex. $w_0 = \lambda$ where $\lambda = 729~\text{nm}$. All values of $|m|$ are within the upper-bound of the topological charge concentration allowed for a beam with this waist size and the waist size itself is achievable with a high NA lens. Note that the peak value decreases with increasing $|m|$ showing no `superchiral' behaviour.}
    \label{fig:1}
\end{figure*}

\begin{figure*}
    \includegraphics[]{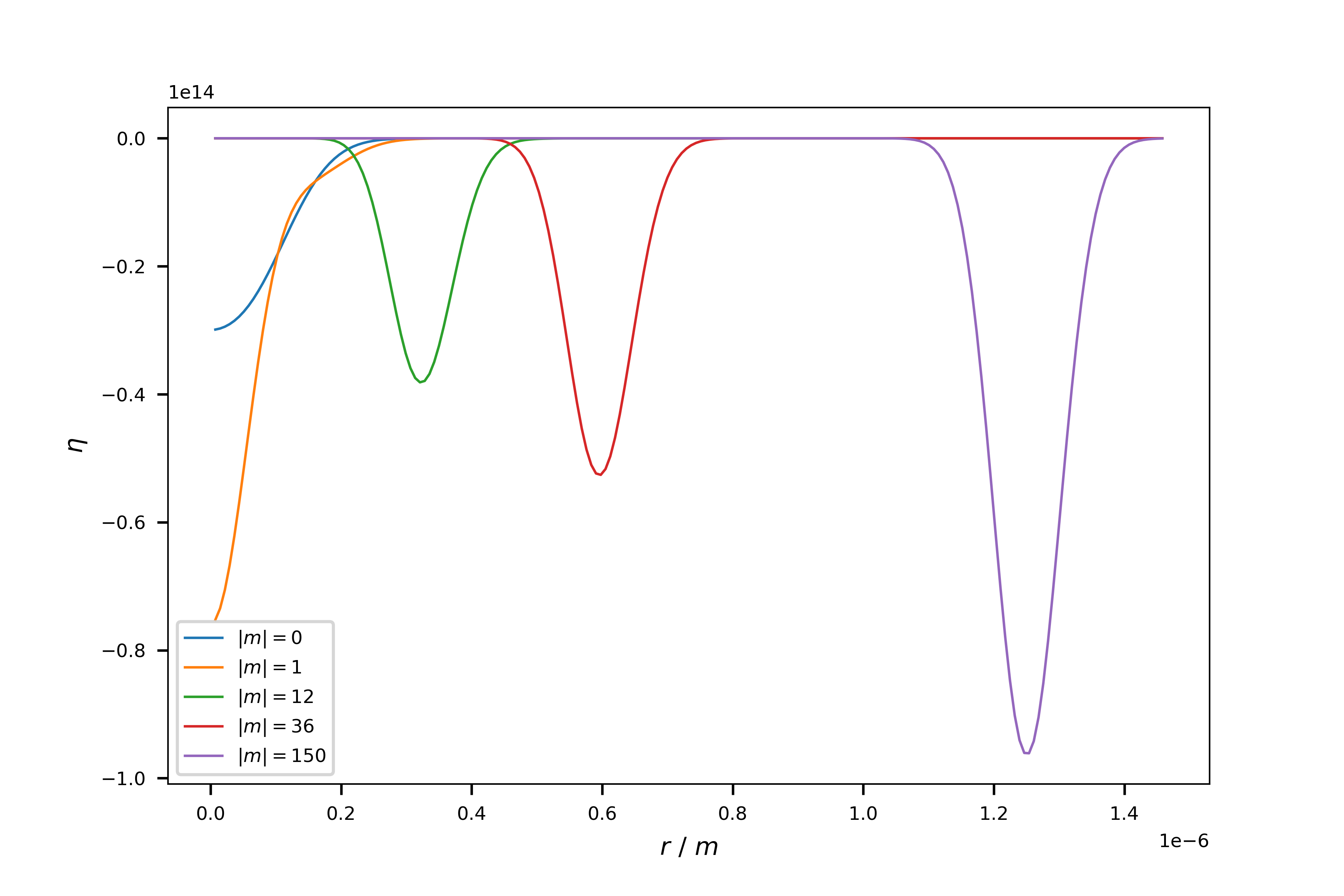}
    \caption{Optical chirality density \eqref{eq:15} for a Laguerre-Gaussian mode $(m,p=0)$ and $\cos \Theta = 0$, i.e. a right-circularly polarized scalar Laguerre-Gaussian vortex. Here we have attempted to reproduce Figure 2 of BYKL (note the sign is flipped because we have used the correct-signed optical chirality density Eq.~\eqref{eq:15}). In BYKL the values of $w_0$ and $\lambda$ used to produce Figure 2 were not given. Here we have used $w_0 \approx \lambda/5$ with $\lambda = 729~\text{nm}$ to yield a decent reproduction of Figure 2 of BYKL. The upper bound on topological charge density however means that under such conditions only values $|m|<2$ are physical; even so, a beam spot size of $w_0 = \lambda /5$ is not physically possible, rendering `superchirality' of these beams nonphysical.}
    \label{fig:2}
\end{figure*}

\section{Further comments}

BYKL makes the statement `\textit{This can be achieved even for moderately focused beams because its contribution to the longitudinal terms is independent of the contribution from tight focusing.}' However, the magnitude of any longitudinal field component of any electromagnetic field is always correlated to the spatial confinement (focusing for laser beams) of the field. This fundamental behaviour is encoded in Heisenberg's uncertainty principle. Let us take a transverse ($x$,$y$) polarized $z$-propagating photon, such that $k\approx k_z$ and thus the magnitude of photon momentum is $p \approx p_z$. The uncertainty principle tells us that the uncertainty in photon momentum in the transverse direction is $\Delta p_x \Delta x \ge \hbar/2$ (the same applies to the $y$-component). This can be re-written and arranged as: $\Delta k_x \ge 1/2\Delta x$ which tells us that unless the photon is spatially confined to wavelength dimensions then the uncertainty in $\Delta k_x$ and the magnitude of the associated $z$-polarized electromagnetic fields is essentially zero. This is why the physical manifestations of longitudinal fields (the transverse spin momentum of light \cite{bliokh2015transverse, aiello2015transverse}, for example) are observed in nano-optics and not paraxial optics \cite{novotny2012principles, adams2018optics, chekhova2021polarization}.

In calculating the optical chirality density BYKL takes the inner product of the electric and magnetic fields which, using the language first introduced by Lax et al. \cite{lax1975maxwell}, include the zeroth-order transverse field $\text{T}_0$ and first-order longitudinal field $\text{L}_1$. The order corresponds to the proportionality of each field component to a smallness parameter. For Gaussian-type beams this factor is $1/kw_0$. In BYKL the calculation is

\begin{align}
\eta &\propto -\text{Im}(\mathbf{E}^* \cdot \mathbf{B}) \nonumber \\
&= -\text{Im}((\mathbf{E_{\text{T0} }^*+E_{\text{L1}}^* })\cdot (\mathbf{B_{\text{T0} }+B_{\text{L1}}})) \nonumber \\
&= -\text{Im}((\mathbf{E_{\text{T0} }^*\cdot{B_{\text{T0}}})+ (\mathbf{E_{\text{L1}}^* }+B_{\text{L1}}})),
\label{eq:16}
\end{align}

where the first term in the bottom line stemming from the zeroth-order transverse fields is zeroth-order in the smallness parameter; the second term proportional to the inner product of the two longitudinal fields, each first-order in the smallness parameter, is thus second-order in the smallness parameter. In BYKL the second-order transverse fields $\mathbf{E_{\text{T2}}}$ and $\mathbf{B_{\text{T2}}}$ has been neglected, which are of course second-order in the smallness parameter. In neglecting these higher-order fields, however, BYKL has also not taken account of the cross terms between the zeroth-order transverse and second-order transverse fields, $\mathbf{E_{\text{T0}}}$, $\mathbf{B_{\text{T0}}}$ and $\mathbf{E_{\text{T2}}}$, $\mathbf{B_{\text{T2}}}$, i.e $\mathbf{E_{\text{T0}}}^*\cdot\mathbf{B_{\text{T2}}}$ and $\mathbf{E_{\text{T2}}}^*\cdot\mathbf{B_{\text{T0}}}$.   These contributions to the optical chirality are of the same order in the smallness parameter as the pure longitudinal fields. For scalar vortex beams it has been highlighted that for linearly-polarized fields these cross-terms are zero; for fields with ellipticity they generate circularly symmetric quantitative corrections \cite{forbes2021measures, forbes2023customized}. However, for vector vortex beams they generate in general very complex patterns which qualitatively and quantitatively significantly influence the optical chirality density and thus cannot be ignored \cite{Forbes2024optical}. Furthermore, these contributions also depend on the Poincar\'e angle $\Phi$. The interested reader is referred to \cite{Forbes2024optical} for further information.  

\section{Conclusion}

The aim of this Comment was to establish that the claims of `superchirality' in Babiker et. al. \cite{babiker2024super} are nonphysical. In order to reproduce the results of \cite{babiker2024super} we had to use unrealistic and unachievable values for the parameters of beam waist $w_0$ and topological charge $m$ (see Fig.~\ref{fig:2}). In this Comment we have highlighted that for experimentally viable values of $w_0$ and $m$, scalar circularly-polarized vortex beams possess no `superchirality' (see Fig.~\ref{fig:1}) and that circular vector beams always possess less optical chirality than the corresponding scalar circularly polarized vortex mode.

\section{Declaration of competing interest}

The author declares that they have no known competing financial interests or personal relationships that could have appeared to influence the work reported in this paper.

\bibliography{references.bib}
\end{document}